\documentclass[preprint,preprintnumbers,amsmath,amssymb]{revtex4}

\usepackage{graphicx}

\usepackage[active]{srcltx}

\usepackage[utf8]{inputenc}

\usepackage{amsfonts}    
\usepackage{amssymb}
\usepackage{latexsym}
\usepackage{rotating}
\usepackage{multirow}

\usepackage{hyperref}
\usepackage{color}

\newcommand{\al}{\alpha}

\newcommand{\rar}{\rightarrow}

\begin{document}

\title{From quartic anharmonic oscillator to double well potential}

\date{\today}

\author{Alexander~V.~Turbiner}
\email{turbiner@nucleares.unam.mx}
\author{J.C.~del~Valle}
\email{delvalle@correo.nucleares.unam.mx}
\affiliation{Instituto de Ciencias Nucleares, Universidad Nacional
Aut\'onoma de M\'exico, Apartado Postal 70-543, 04510 M\'exico,
D.F., Mexico{}}

\begin{abstract}
It is already known that the quantum quartic single-well anharmonic oscillator $V_{ao}(x)=x^2+g^2 x^4$ and double-well anharmonic oscillator $V_{dw}(x)= x^2(1 - gx)^2$ are essentially one-parametric, their eigenstates depend on a combination $(g^2 \hbar)$. Hence, these problems are reduced to study the potentials $V_{ao}=u^2+u^4$ and $V_{dw}=u^2(1-u)^2$, respectively. It is shown that by taking uniformly-accurate approximation for anharmonic oscillator eigenfunction $\Psi_{ao}(u)$, obtained recently, see JPA 54 (2021) 295204 [1] and Arxiv 2102.04623 [2], and then forming the function $\Psi_{dw}(u)=\Psi_{ao}(u) \pm \Psi_{ao}(u-1)$ allows to get the highly accurate approximation for both the eigenfunctions of the double-well potential and its eigenvalues.
\end{abstract}


\maketitle

\section{Introduction}
For the one-dimensional quantum quartic single-well anharmonic oscillator $V_{ao}(x)=x^2+g^2 x^4$ and double-well anharmonic oscillator with potential $V_{dw}(x)= x^2(1 - gx)^2$ the (trans)series in the coupling constant $g$ (which is the Perturbation Theory in powers of $g$ (the Taylor expansion) in the former case of $V_{ao}(x)$ supplemented by exponentially-small terms in $g$ in the latter case of $V_{dw}(x)$) and the semiclassical expansion in $\hbar$ (the Taylor expansion for $V_{ao}(x)$ supplemented by the exponentially small terms in $\hbar$ for $V_{dw}(x)$) for energies {\it coincide} \cite{Shu-Tur:2018}. This property plays crucially important role in our consideration.

Both the quartic anharmonic oscillator
\begin{equation}
\label{AHO}
  V\ =\ x^2 + g^2 x^4 \ ,
\end{equation}
with a single harmonic well at $x=0$ and the double-well potential
\begin{equation}
\label{DW}
  V\ =\ x^2(1 - g x)^2 \ ,
\end{equation}
with two symmetric harmonic wells at $x=0$ and $x=1/g$, respectively, are particular cases
of the quartic polynomial potential
\begin{equation}
\label{PAHO}
  V\ =\ x^2 + a g x^3 + g^2 x^4 \ ,
\end{equation}
where $g$ is the coupling constant and $a$ is a parameter. Interestingly, the potential (\ref{PAHO}) is symmetric for three particular values of the parameter $a$:
$a=0$ and $a=\pm 2$. All three potentials (\ref{AHO}), (\ref{DW}), (\ref{PAHO}) belong to the family of potentials of the form
\[
    V\ =\ \frac{1}{g^2}\ {\tilde V} (gx) \ ,
\]
for which there exists a remarkable property: the Schr\"odinger equation becomes one-parametric, both the Planck constant $\hbar$ and the coupling constant $g$ appear in the combination $(\hbar g^2)$, see \cite{Shu-Tur:2021}. It can be immediately seen if instead of the coordinate $x$ the so-called classical coordinate $u=(g\,x)$ is introduced. This property implies that the action $S$ in the path integral formalism becomes $g$-independent and the factor $\frac{1}{\hbar}$ in the exponent becomes $\frac{1}{\hbar g^2}$ \cite{EST:2016}. Formally, the potentials (\ref{AHO})-(\ref{DW}), which enter to the action, appear at $g=1$, hence, in the form
\begin{equation}
\label{AHO-u}
  V\ =\ u^2 + u^4 \ ,
\end{equation}
\begin{equation}
\label{DW-u}
  V\ =\ u^2(1 - u)^2 \ ,
\end{equation}
respectively. Both potentials are symmetric with respect to $u=0$ and $u=1/2$, respectively.

Namely, this form of the potentials will be used in this short Note. This Note is the extended version of a part of presentation in AAMP-18 given by the first author \cite{AHO-Prague}.

\section{Single-well potential}
\label{sect:single}

In \cite{AHO} for the potential (\ref{AHO-u}) matching the small distances $u \rar 0$ expansion and the large distances $u \rar \infty$ expansion (in the form of semiclassical expansion) for the phase $\phi$ in the representation
\[
     \Psi\ =\ P(u)\ e^{-\phi(u)} \ ,
\]
of the wave function, where $P$ is a polynomial, it was constructed the following function for the $(2n+p)$-excited state with quantum numbers $(n,p)$, $n=0,1,2,\ldots\ ,\ p=0,1$\,:
\[
 \Psi^{(n,p)}_{(approximation)}\ =\
\]
\[
 \frac{u^p P_{n,p}(u^2)}{\left(B^2\ +\ u^2 \right)^{\frac{1}{4}}
 \left({B}\ +\ \sqrt{B^2\ +\ u^2} \right)^{2n+p+\frac{1}{2}}}
\]
\begin{equation}
\label{AHO-psi}
   \exp \left(-\ \dfrac{A\ +\ (B^2 + 3)\,u^2/6\ +\ u^4/3}
  {\sqrt{B^2\ +\ u^2}} \ +\ \frac{A}{B}\right)\ ,
\end{equation}
where $P_{n,p}$ is some polynomial of degree $n$ in $u^2$ with positive roots. Here $A=A_{n,p},\ B=B_{n,p}$ are two parameters of interpolation. These parameters $(-A),\ B$ are slow-growing with quantum number $n$ at fixed $p$ taking, in particular, the values
\begin{equation}
\label{AB-00}
A_{0,0} =-0.6244\ , B_{0,0}=2.3667\ ,
\end{equation}
\begin{equation}
\label{AB-01}
A_{0,1} =-1.9289\ , B_{0,1}=2.5598\ ,
\end{equation}
for the ground state and the first excited state, respectively. This remarkably simple function (\ref{AHO-psi}), see Fig.1 (top), provides 10-11 exact figures in energies for the first 100 eigenstates. Furthermore, the function (\ref{AHO-psi}) deviates uniformly for $u \in (-\infty, +\infty)$ from the exact function in $\sim 10^{-6}$.

\begin{figure}
\centering
\includegraphics[width=7cm]{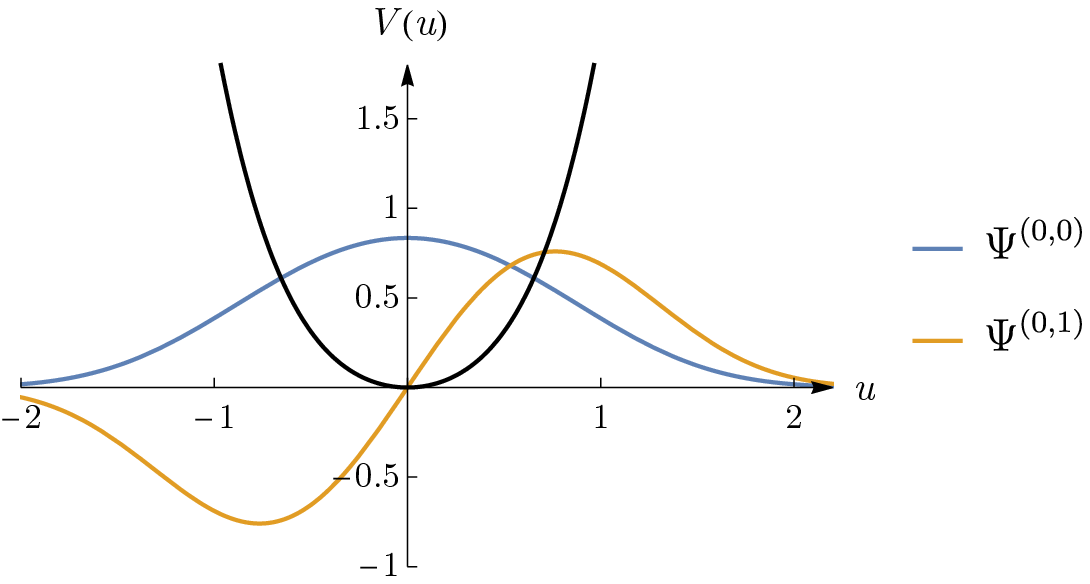} 
\\[5mm]
\includegraphics[width=7cm]{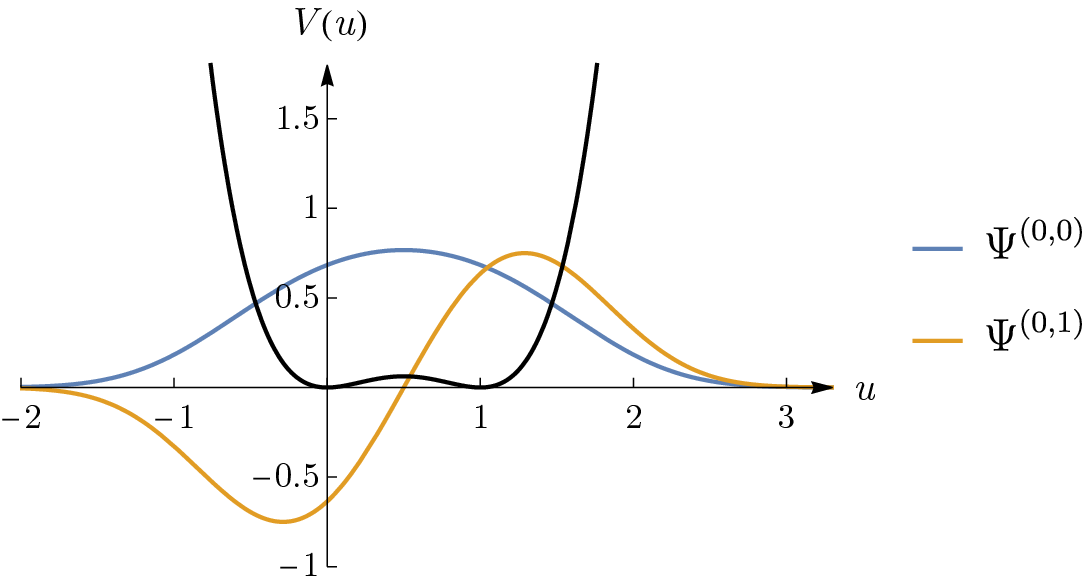} 
\caption{Two lowest, normalized to one eigenfunctions of positive/negative parity:
for single-well potential (\ref{AHO-u}), see (\ref{AHO-psi}) (top) and for double-well potential (\ref{DW-u}), see (\ref{final})(bottom).  Potentials shown by black lines.}
\label{fig:res}
\end{figure}

\section{Double-well potential: wavefunctions}
\label{sect:double}

Following the prescription, usually assigned in folklore to E.M.~Lifschitz - one of the authors of the famous Course on Theoretical Physics by L.D.~Landau and E.M.~Lifschitz - when a wavefunction for single well potential with minimum at $u=0$ is known, $\Psi(u)$, the wavefunction for double well potential with minima at $u=0, 1$ can be written as $\Psi(u) \pm \Psi(u-1)$. This prescription was already checked successfully for the double-well potential (\ref{DW}) in \cite{Turbiner:2010} for somehow simplified version of (\ref{AHO-psi}), based on matching the small distances $u \rar 0$ expansion and the large distances $u \rar \infty$ expansion for the phase $\phi$ but ignoring subtleties emerging in semiclassical expansion.
Taking the wavefunction (\ref{AHO-psi}) one can construct

\[
 \Psi^{(n,p)}_{(approximation)}\ =\
\]
\[
 \frac{P_{n,p}({\tilde u}^2)}{\left(B^2\ +\ {\tilde u}^2 \right)^{\frac{1}{4}}
 \left({\al B}\ +\ \sqrt{B^2\ +\ {\tilde u}^2} \right)^{2n+\frac{1}{2}}}
\]
\begin{equation}
\label{final}
   \exp \left(-\ \dfrac{A\ +\ (B^2 + 3)\,{\tilde u}^2/6\ +\ {\tilde u}^4/3}
  {\sqrt{B^2\ +\ {\tilde u}^2}} \ +\ \frac{A}{B}\right)\ D^{(p)}\ ,
\end{equation}
where $p=0,1$ and
\[
  D^{(0)}\ =\ \cosh \bigg( \frac{a_0 {\tilde u} + b_0 {\tilde u}^3}{\sqrt{B^2 + {\tilde u}^2}} \bigg)\ ,
\]
\[
  D^{(1)}\ =\ \sinh \bigg( \frac{a_1 {\tilde u} + b_1 {\tilde u}^3}{\sqrt{B^2 + {\tilde u}^2}} \bigg) \ .
\]
Here
\begin{equation}
	\tilde{u}\ =\ u\ -\  \frac{1}{2} \ ,
	\end{equation}
$\al=1$ and $A, B, a_{0,1}, b_{0,1}$ are variational parameters. If $\al=0$ as well as $b_{0,1}=0$ the function (\ref{final}) is reduced to ones which were explored in \cite{Turbiner:2010}, see Eqs.(10)-(11). The polynomial $P_{n,p}$ is found unambiguously
after imposing the orthogonality conditions of $ \Psi^{(n,p)}_{(approximation)}$ to
$\Psi^{(k,p)}_{(approximation)}$ at $k=0,1,2,\ldots , (n-1)$, here it is assumed that the polynomials $P_{k,p}$ at $k=0,1,2,\ldots , (n-1)$ are found beforehand.

\section{Double-well potential: Results}
\label{sect:double-R}

In this section we present concrete results for energies of the ground state $(0,0)$ and of the first excited state $(0,1)$ obtained with the function (\ref{final}) at $p=0,1$, respectively.
The results are compared with the Lagrange-Mesh Method (LMM) \cite{Tur-delValle:2021}.

\subsection{Ground State (0,0)}

The ground state energy for (\ref{DW-u}) obtained variationally using the function (\ref{final}) at $p=0$ and compared with LMM results \cite{Tur-delValle:2021}, where all printed digits (in the second line) are correct,
\begin{align*}
	E_{var}^{(0,0)}\ &=\ 0.932\,517\,518\,401 \ , \\
    E_{mesh}^{(0,0)}\ &=\ 0.932\,517\,518\,372 \ .
\end{align*}
Note that ten decimal digits in $E_{var}^{(0,0)}$ coincide with ones in $E_{mesh}^{(0,0)}$ (after rounding). Variational parameters in (\ref{final}) take values,
%
\begin{align*}
A\ &=\   2.3237\ , \\
B\ &=\   3.2734\ , \\
a_0\ &=\  2.3839\ ,  \\ 	
b_0\ &=\  0.0605 \ ,
\end{align*}
cf.(\ref{AB-00}). Note that $b_0$ takes a very small value.

\subsection{First Excited  State (0,1)}

The first excited state energy for (\ref{DW-u}) obtained variationally using the function (\ref{final}) at $p=1$ and compared with LMM results \cite{Tur-delValle:2021}, where all printed digits (in the second line) are correct,
\begin{align*}
	E_{var}^{(0,1)}\ &=\ 3.396\,279\,329\,936 \ , \\
	E_{mesh}^{(0,1)}\ &=\ 3.396\,279\,329\,887 \ .
\end{align*}
Note that ten decimal digits in $E_{var}^{(0,1)}$ coincide with ones in $E_{mesh}^{(0,1)}$ (after rounding). Variational parameters in (\ref{final}) take values,
\begin{align*}
	A\ &=\  -2.2957 \ ,\\
	B\ &=\   3.6991 \ ,\\
	a_1\ &=\   4.7096 \ ,\\ 	
	b_1\ &=\   0.0590 \ ,
\end{align*}
cf.(\ref{AB-01}). Note that $b_1$ takes a very small value similar to $b_0$.

\section{Conclusions}

It is presented the approximate expression (\ref{final}) for the eigenfunctions in the double-well potential (\ref{DW-u}). In Non-Linearization procedure \cite{T:1984} it can be calculated the first correction (the first order deviation) to the function (\ref{final}). It can be shown that for any $u \in (-\infty, +\infty)$ the functions (\ref{final}) deviate uniformly from the exact eigenfunctions, beyond the sixth significant figure similarly to the function (\ref{AHO-psi}) for the single-well case. It increases the accuracy of the simplified function, proposed in [5] with $\al=0$ and $b_{0,1}=0$, in the domain under the barrier $u \in (0.25 , 0.75)$ from 4 to 6 significant figures leaving the accuracy outside of this domain practically unchanged.

\begin{acknowledgements}
This work is partially supported by CONACyT grant A1-S-17364 and DGAPA grant IN113819 (Mexico).
AVT thanks the PASPA-UNAM program for support during his sabbatical leave.
\end{acknowledgements}


\end{document}